\documentclass[useAMS,usenatbib]{mn2e}

\usepackage{lscape}
\usepackage{supertabular}
\usepackage{longtable}
\usepackage{graphicx}
\usepackage{txfonts}
\def\gtsima{$\; \buildrel > \over \sim \;$}
\def\ltsima{$\; \buildrel < \over \sim \;$}
\def\gsim{\lower.5ex\hbox{\gtsima}}
\def\lsim{\lower.5ex\hbox{\ltsima}}

\title[Time lags in NGC 5408 X-1]     
  {Time lags in the ultraluminous X-ray source NGC 5408 X-1: implications for the black hole mass}         
\author[B. De Marco et al]
  {B.~De Marco,$^{1,2}$\thanks{E-mail: bdemarco@mpe.mpg.de} 
  G.~Ponti,$^2$ G.~Miniutti,$^1$
  T.~Belloni,$^3$  M.~Cappi,$^4$, M.~Dadina$^4$, T.~Mu\~noz-Darias$^5$\\ 
  $^1$Centro de Astrobiolog\'{i}a (CSIC-INTA), Dep. de Astrof\'{i}sica; ESAC, PO Box 78, E-28691, Villanueva de la Ca\~nada, Madrid, Spain\\
  $^2$Max-Planck-Institut f\"{u}r extraterrestrische Physik, Giessenbachstrasse 1, D-85748 Garching bei M\"{u}nchen, Germany\\
  $^3$INAF-Osservatorio Astronomico di Brera, Via E. Bianchi 46, 23807 Merate (LC), Italy\\
  $^4$INAF-IASF Bologna, Via Gobetti 101, 40129 Bologna, Italy\\
  $^5$School of Physics and Astronomy, University of Southampton, Southampton, Hampshire, SO17 1BJ, United Kingdom\\
}
\date{Released 2012 Xxxxx XX}

\pagerange{\pageref{firstpage}--\pageref{lastpage}} \pubyear{2012}

\def\LaTeX{L\kern-.36em\raise.3ex\hbox{a}\kern-.15em
    T\kern-.1667em\lower.7ex\hbox{E}\kern-.125emX}

\begin{document}

\label{firstpage}

\maketitle

\begin{abstract}
We present the analysis of the X-ray variability and spectral timing properties of the ultraluminous X-ray source (ULX) NGC 5408 X-1, one of the most variable ULXs known so far. 
The variability properties are used as a diagnostic of the accretion state of the source and to derive estimates of the black hole (BH) mass. The observed high level of fast X-ray variability (fractional root-mean-square variability -- \emph{rms} -- amplitude of $\sim 30$ percent in the hard energy band), the hardening of the fractional \emph{rms} spectrum, and the properties of the QPO, all resemble those of a source in a hard-intermediate accretion state. 
We confirm the previous detection of a soft lag in the X-ray light curves of the source during 2006 and 2008 observations and find that the soft lag is still present in the more recent 2010/2011 observations. Using the entire available XMM-Newton data set (public as of February 2012) we observe that the soft lag (of few seconds amplitude) is detected over a relatively large range of frequencies ($\nu \sim 5-90$ mHz), which always includes the QPO frequencies. The soft lag displays energy-dependence, with the (absolute) amplitude increasing as a function of energy separation.
We find close analogies with soft lags associated to type-C QPOs in BH binary systems (although an association to other types of QPOs cannot be completely excluded), as well as with reverberation lags observed in AGN. In both cases an intermediate mass BH (IMBH) solution appears the most plausible.

\end{abstract}


\section{Introduction}
Ultraluminous X-ray sources (ULX, e.g. Fabbiano 2004, Roberts 2007) are off-nuclear, point-like sources, 
with luminosity exceeding $L_X\sim 10^{39}$erg s$^{-1}$.
Timing and spectral variability studies yielded sufficient evidence of the majority of them being 
 accreting compact objects. 
However, their high luminosity is not easy to reconcile with emission from standard black hole binary (BHB) systems. 
Indeed, assuming isotropic emission, the observed X-ray fluxes require either super-Eddington accretion onto a stellar mass black hole (M$_{BH}\sim 10$M$_{\odot}$), 
or sub-Eddington accretion onto an intermediate mass black hole (IMBH, i.e. M$_{BH}\sim 10^{2-4}$M$_{\odot}$, 
e.g. Colbert \& Mushotzky 1999). Alternative explanations imply the enhanced brightness to be due to beamed emission (King et al 2001). 
Overall, a combination of mild beaming and moderately super-Eddington rate seems to account for the majority of the observed ULXs up to luminosities of $L_X\sim 10^{40}$erg s$^{-1}$. However, the IMBH interpretation remains a plausible explanation for the most luminous sources (Feng \& Soria 2011 and references therein).
 Unfortunately, it has not been possible so far to 
rely on direct mass estimates, due to the difficulties in detecting the companion star.
Thus, the most promising way to deal with the ULX issue is to make use of indirect methods, e.g. by comparing their spectral and 
timing properties with those of well known accreting X-ray sources (i.e. BHBs, AGNs, e.g. Casella et al 2008).\\
Mass-scaling arguments suggest that the same mechanism can explain the observed spectral and timing properties of accreting objects spanning up to $\sim 8$ orders of magnitude in mass (from neutron stars to AGN, e.g. Merloni, Heinz, \& di Matteo 2003; McHardy et al 2006; Koerding et al 2007). According to this scenario ULXs are expected to fit in this picture as well, with their specific properties (e.g. characteristic time scales, disc temperature) depending on the BH mass. 
One of the main observables characterizing X-ray binaries is their accretion state. X-ray binaries are generally in a quiescent state, occasionally broken up by violent outbursts, during which the spectral and timing properties of the source undergo major changes over relatively short time scales (Belloni 2010, Belloni et al 2011). At the beginning of the outburst the source is in the power law-dominated, low hard state (LHS). This state spans a large range of X-ray luminosities, and is characterized by a high X-ray variability power and the presence of a jet (Markoff et al 2001, Fender et al 2004). When the source leaves the hard state it gradually softens, up to the so called disc-dominated, high soft state (HSS), during which the X-ray variability power strongly decreases, while the jet is quenched, and the emergence of disc winds is observed (Ponti et al 2012). In between these two canonical states the source passes through the hard-intermediate (HIMS) and soft-intermediate (SIMS) states, namely transition states mainly characterized by a gradual increase of disc flux fraction, with a corresponding gradual decrease of variability power, and the presence of QPOs.
Koerding et al (2006) demonstrated that the concept of accretion states can be applied to AGN as well, once the proper scaling to longer time scales is taken into account.\\
It has been observed that the total amount of variability strongly depends on the accretion state of the source and can be used as a good tracer of it (Mu\~noz-Darias et al 2011).
NGC 5408 X-1 is a well known ULX, showing remarkable X-ray variability (Heil et al 2009). Moreover, it is one of the few ULXs (e.g. M82 X-1, Strohmayer \& Mushotzky 2003, Mucciarelli 
et al 2006) for which the detection of a low-frequency quasi-periodic oscillation (QPO) has been reported (Strohmayer et al 2007, Strohmayer \& Mushotszky 2009, hereafter S07 and SM09 respectively).
The properties of the detected QPO in NGC 5408 X-1 favour an association with the type C QPOs (S07) commonly 
observed in BHBs during LHS and HIMS (e.g. Casella et al 2004). Such association, combined with simple frequency-scaling arguments, implies NGC 5408 X-1 to host an 
IMBH with M$\sim 10^{2-3}$M$_{\odot}$ (e.g.  Mucciarelli et al 2006, Casella et al 2008). However, this interpretation is not univocal, 
and different scenarios have been proposed (e.g. a M$_{BH}\leq 100$ M$_{\odot}$ accreting at super-Eddington rate, Middleton et al 2011).\\
In this paper we present a comprehensive analysis of the variability and spectral timing properties of NGC 5408 X-1, and use them as diagnostics of the accretion state of the source. Moreover, we report the detailed study of the soft X-ray time-lag in NGC 5408 X-1 light curves. We discuss the analogies with soft X-ray lags observed in other accreting BH systems, and derive implications on the BH mass.

\section{Observations and data reduction}

We analyzed the six archived (see Table \ref{log}) XMM-Newton observations of NGC 5408 X-1, carried out between January 2006 and January 2011. 
\begin{table}
\caption{Observation log: ID number, satellite revolution number, date, effective exposure of combined EPIC pn$+$MOS1$+$MOS2 data after selection of good time intervals, and bibliographic references for each data set.}
\label{log}
\centering
\vspace{0.2cm}
\begin{scriptsize}
\begin{tabular}{p{0.99cm}p{0.99cm}p{0.99cm}p{0.99cm}p{0.99cm}}
\hline\hline  
Obs ID & Rev & date & exp      & ref \\
           &                &         & (ks)       &   \\
\hline
0302900101 & 1117 & 2006-01-13 &  90 & S07 \\
0500750101 & 1483 & 2008-01-13 & 53 & SM09\\
0653380201 & 1942 & 2010-07-17 & 90 & PS12 \\
0653380301 & 1943 & 2010-07-19 & 112 & PS12 \\
0653380401 & 2039 & 2011-01-26 & 92 & PS12 \\
0653380501 & 2040 & 2011-01-28 & 99 & PS12 \\
\hline
\hline
\end{tabular}
\end{scriptsize}
\end{table}
We used combined data from EPIC pn$+$MOS1$+$MOS2 detectors, so as to get a higher signal-to-noise ratio (S/N).
Data reduction is carried out using XMM Science Analysis System (SAS v. 10.0), following standard procedures. 
Time intervals characterized by high background proton flares were filtered out. The selected good time intervals (GTIs) are those during which 
all the three instruments are simultaneously free from background proton flares. This selection reduced 
the useful exposure to a total of $\sim$540 ks. 
Source counts were extracted from a 40 arcsec radius region, while rectangular regions, on the same chip of the source and avoiding other sources, were used for the background.
Only events with PATTERN $\leq 4$  and PATTERN $\leq 12$ were selected respectively for EPIC pn and MOS data. 
The same start and stop time have been chosen for the time series obtained from the different detectors.
We extracted spectra grouping them to a minimum of 20 counts per bin and building responce matrices through the 
RMFGEN and ARFGEN SAS tasks. Using XSpec software package v. 12.5.1, we verified that above 7 keV the S/N is very low, being the background-subtracted source counts of the same order of the background counts. Given this, we decided to restrict our timing analysis to energies E$\leq7$keV.\\

\section{Analysis}
\label{sect:bands}
The XMM-Newton energy spectra of NGC 5408 X-1 reveal the presence of two distinctive X-ray features: an excess of emission at E$\lsim1$keV above the extrapolation of the best fit high-energy power law, and a break or turnover at E$\sim 6$ keV (for detailed spectral analysis see e.g. Gladstone et al 2009, Caballero-Garc\'{i}a \& Fabian 2010; Middleton et al 2011; Pasham \& Strohmayer 2012, hereafter PS12). 
The same features characterize the spectrum of other ULXs and their physical origin is debated (e.g. Walton et al 2011, Feng \& Soria 2011 and references therein).
Despite the intrinsic differences, all proposed models that seek to explain the observed properties of NGC 5408 X-1 and other ULXs with a similar X-ray spectrum (e.g. Walton et al 2011), require the presence of two distinct spectral components in the soft and hard bands. Thus, for the timing analysis of NGC 5408 X-1, we chose to define the two bands 0.3-1 keV and 1-7 keV that select, respectively, the soft and hard main spectral components of the X-ray spectrum of the source (see Gladstone et al 2009, Caballero-Garc\'{i}a \& Fabian 2010, Middleton et al 2011).

\begin{figure} 
\centering

\begin{tabular}{p{4.4cm}}
\includegraphics[height=5.5cm,angle=270]{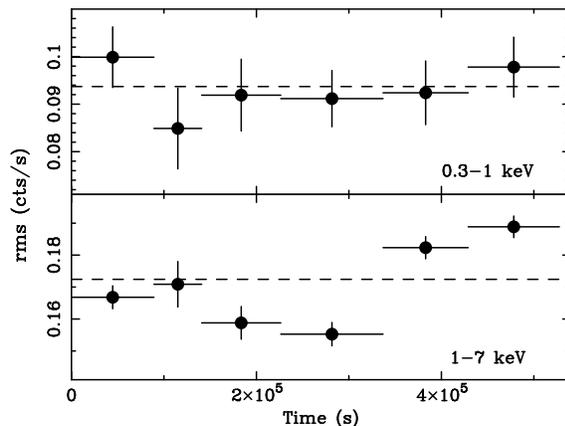} 
\end{tabular}
\caption{The 1-50 mHz \emph{rms} as computed in the soft (0.3-1 keV, \emph{upper panel}) and hard (1-7 keV, \emph{lower panel}) energy bands for the different observations. The dashed horizontal lines represent the best fit constant model.}
\label{fig:stat}

\end{figure}

\subsection{Stationarity}
\label{sect:stat}
\begin{figure} 
\centering

\begin{tabular}{p{4.4cm}}
\includegraphics[height=5.5cm,angle=270]{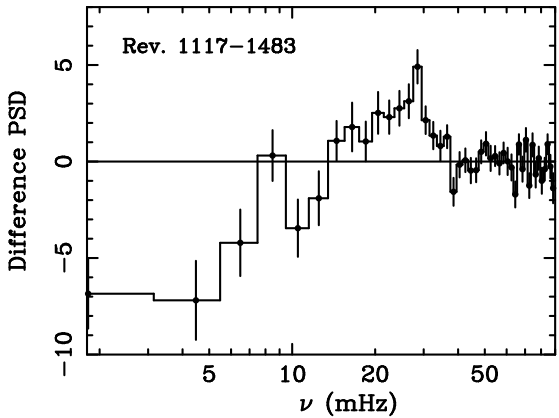} 
\end{tabular}
\caption{Difference between the PSDs of Rev. 1117 and 1483 in the hard 1-7 keV energy band.}
\label{fig:PSDdiff}

\end{figure}

The analyzed dataset covers a time interval of 5 years, with consecutive observations separated by up to $\sim$ 1-2 years. We verified whether the short-time scale variability of the source is consistent with being stationary among the different monitorings. To this aim we followed the method described in Vaughan et al (2003), and computed the average root-mean-square variability amplitude, \emph{rms}, to trace possible deviations from stationarity. For each observation, we extracted light curves in the soft, 0.3-1 keV, and hard, 1-7 keV, energy ranges at 10 s time resolution, and divided them into bits of 1000 s (note that each effective exposure is long enough to ensure a number of segments $>>20$ per observation, see Table \ref{log}). From every light curve segment we derived an estimate of the \emph{rms}, by integrating the power spectrum (PSD) over all the sampled frequencies, i.e. $1-50$ mHz. Averaging these \emph{rms} values we obtained a robust estimate of the \emph{rms} of each observation for the soft and hard energy bands, as well as of the associated standard deviation (which accounts for both Poisson noise and red noise statistical fluctuations). Results are shown in Fig. \ref{fig:stat}. A fit with a constant yields $\chi^2_{\nu}=0.54$ and $\chi^2_{\nu}=12.7$, respectively for the soft and hard band, revealing strong non-stationarity in the hard band. The larger relative errors in the soft band do not allow to draw similar conclusions.\\
In order to cross-check these results, we directly inspected the 1-7 keV broad band power spectra, and tested for intrinsic variations in their shape, by computing the difference between the PSDs of consecutive observations. To quantify the significance level of deviations from stationarity we carried out a $\chi^2$ goodness-of-fit test against a constant model fixed at the value of zero, in the same frequency range as the one used for the \emph{rms}, i.e. $\nu=1-50$ mHz. We set at 0.1 percent the confidence level for rejection of our null hypothesis (namely that a constant model fixed at zero provides a good description of the difference PSD). This limit corresponds to a probability of $>3\sigma$ for the source to be non-stationary between consecutive observations.
According to this test, stationarity is observed between Revs. 1942 and 1943, and between Revs. 2039 and 2040, while other observations are inconsistent with being stationary at $>4 \sigma$. Overall these results are in agreement with those obtained from the comparison of \emph{rms} estimates, with the exception of Revs. 1117 and 1483, which have similar \emph{rms} but the PSD shapes differ, with Rev. 1117 characterized by lower variability power at low frequencies, and excess power between 15-40 mHz (see Fig. \ref{fig:PSDdiff}).

\subsection{Accretion state from variability}
\label{sect:rms}
\begin{figure} 
\centering

\begin{tabular}{p{4.4cm}}
\includegraphics[height=5.5cm,angle=270]{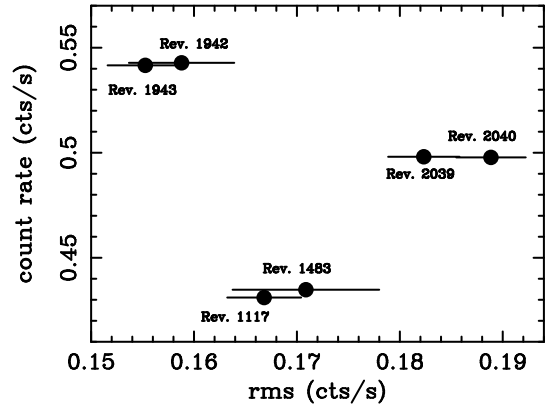} 
\end{tabular}
\caption{The hard band (1-7 keV), 1-50 mHz\emph{rms} against average hard band count rate for the different observations.}
\label{fig:rms_cr}

\end{figure}

\begin{figure} 
\centering

\begin{tabular}{p{4.4cm}}
\includegraphics[height=5.8cm,angle=270]{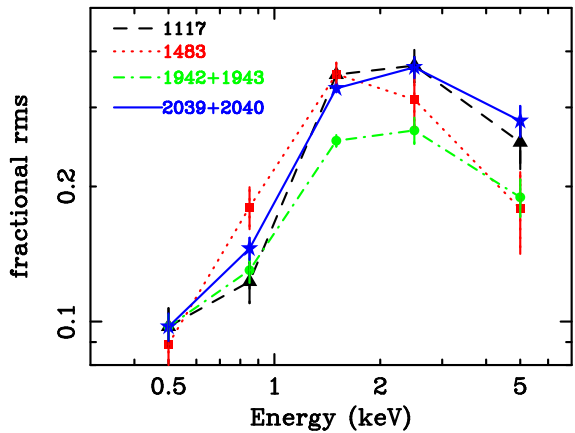} 
\end{tabular}
\caption{The fractional \emph{rms} spectrum in the frequency interval 1-50 mHz.}
\label{fig:rms_spec}

\end{figure}

Deviations from stationarity may be a consequence of correlated \emph{rms} and source flux variations (e.g. Vaughan et al 2003). Mu\~noz-Darias et al (2011) showed that the evolution of the short time scale variability as a function of flux is a good tracer of the accretion regime in BH X-ray binaries. A linear relation of \emph{rms} with count rate characterized by a very small scatter is observed only during the canonical LHS, while a much more scattered linear relation identifies soft states. On the other hand, a more chaotic behaviour, mostly characterized by a decrease of \emph{rms} as the flux increases, is observed during intermediate states.\\ 
Mass-scaling arguments suggest that all accreting objects should undergo similar state transitions, on a time scale that depends on the mass of the system (e.g. see Koerding et al 2006 for analogies between AGN and X-ray binaries). Evidences of state transition have been recently reported also for a ULX (XMMU J004243.6$+$412519,  Middleton et al 2013), while analogies with canonical states of X-ray binaries can be found in many ULXs (see Feng \& Soria 2011 and references therein). It is thus interesting to check whether the variability properties of NGC 5408 X-1 can be used to classify its XMM-Newton observations in the context of known X-ray binaries accretion states.\\
Following results by Mu\~noz-Darias et al (2011), we plot in Fig. \ref{fig:rms_cr} the hard band \emph{rms} as a function of the average hard band count rate. 
 The absence of a linear correlation is clear from the data. This excludes NGC 5408 X-1 to be observed in an accretion state similar to the canonical LHS of BH X-ray binaries.
It is worth noting that this conclusion agrees with results from independent spectral analyses, which show the presence of a strong soft component in the data (see Sect. \ref{sect:bands}).\\
We studied the variability of the source as a function of energy by deriving the fractional \emph{rms} spectrum (e.g. Vaughan et al 2003) of each observation in the $1-50$ mHz frequency interval. According to results presented in Sect. \ref{sect:stat}, the variability properties of the source do not change significantly between revolutions adjacent in time (see also Figs. \ref{fig:stat} and \ref{fig:rms_cr}). For this reason consecutive observations during which the source appears stationary have been taken as different realizations of the same process and combined (i.e. Revs. $1942+1943$, and Revs. $2039+2040$) by averaging their PSDs. The fractional \emph{rms} spectrum of the three datasets is shown in Fig. \ref{fig:rms_spec}. The \emph{rms} and errors are obtained following the same procedure described in Sect. \ref{sect:stat}, and normalizing the \emph{rms} for the mean count rate.
The variability of the source is observed to increase with energy (in agreement with results by Middleton et al 2011 for Revs. 1117 and 1483), up to $\sim$3 keV, where it appears to drop to lower values (with a deviation of $>3\sigma$ from the extrapolation of a constant fit to the 1-3 keV plateau). 
The soft band is characterized by lower, but still significant, levels of variability ($\sim$10 percent), while in the hard band it increases to $\sim$30 percent. 
Revs. 1942+1943 show a significant decrease of hard band variability.\\
The observed hardening of the fractional \emph{rms} is reminiscent of the spectral shape characterizing the fractional \emph{rms} of X-ray binaries during the softest stages of HIMS (i.e. just before the transition), during SIMS, and/or during HSS (Belloni et al 2011). However, we note that BHBs are characterized by very low levels of variability during disc-dominated soft states (fractional \emph{rms} of few percent). This is at odds with the high fractional variability registered for NGC 5408 X-1, thus disfavouring the canonical HSS interpretation, but consistent with the source being catched during an intermediate state.

\subsection{Power spectral shape and QPOs}
\label{Sect:PSD}
 
\begin{figure*} 
\centering
\begin{tabular}{p{8.cm}p{8.cm}}
\includegraphics[height=8.cm,width=6.8cm,angle=0]{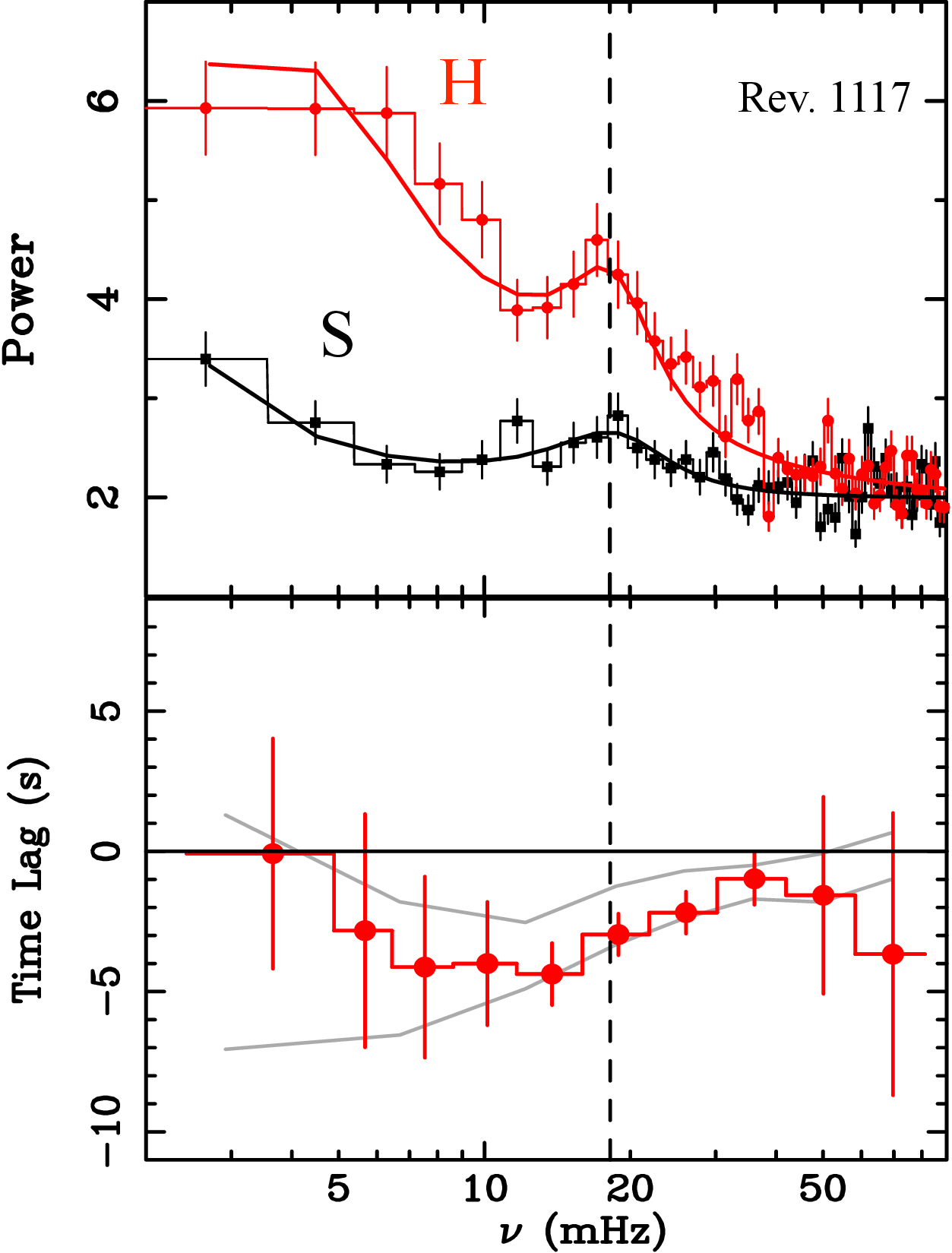} & \includegraphics[height=8.cm,width=6.8cm,angle=0]{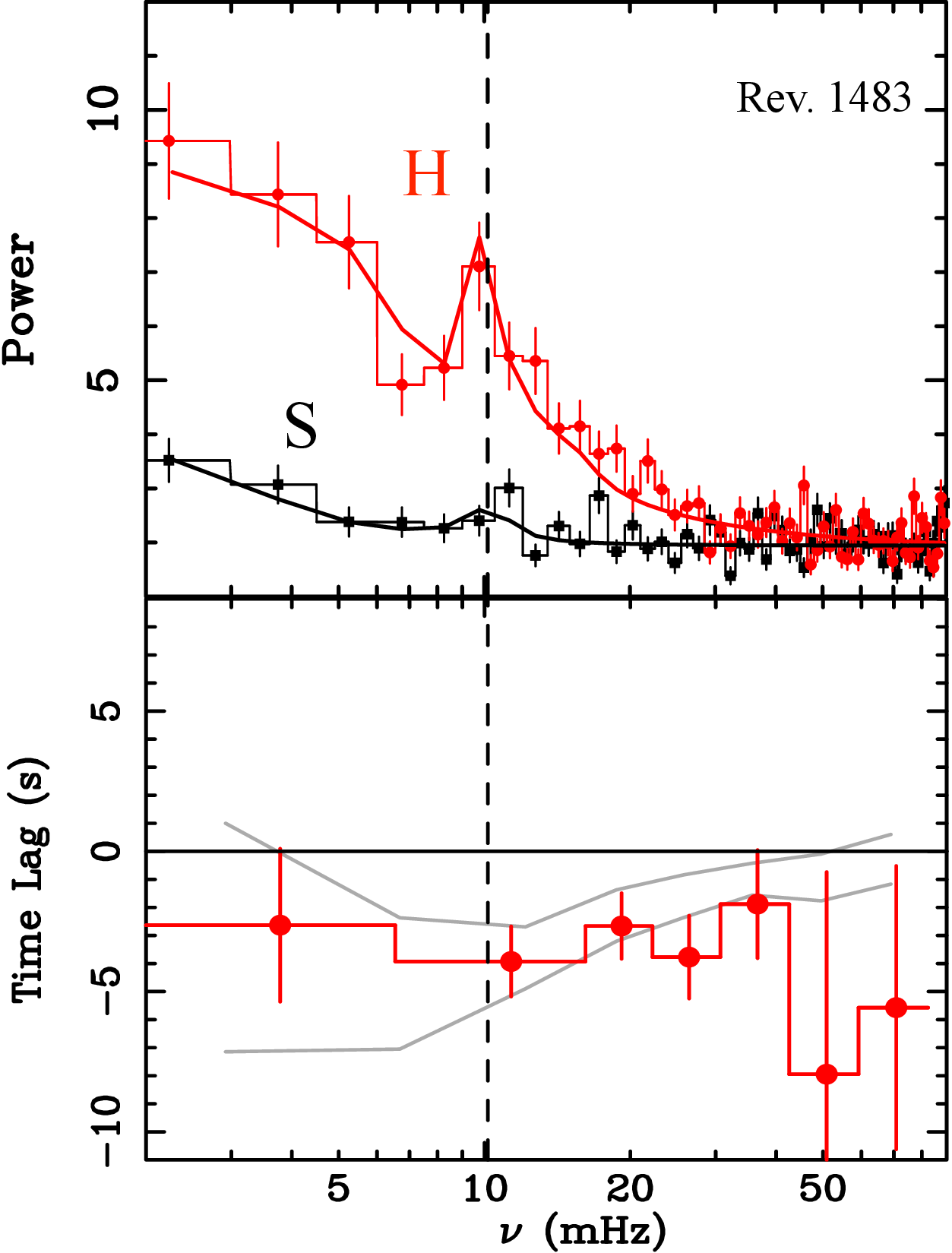}\\ 
\end{tabular}
\vspace{1.cm}
\begin{tabular}{p{8.cm}p{8.cm}}
\includegraphics[height=8.cm,width=6.8cm,angle=0]{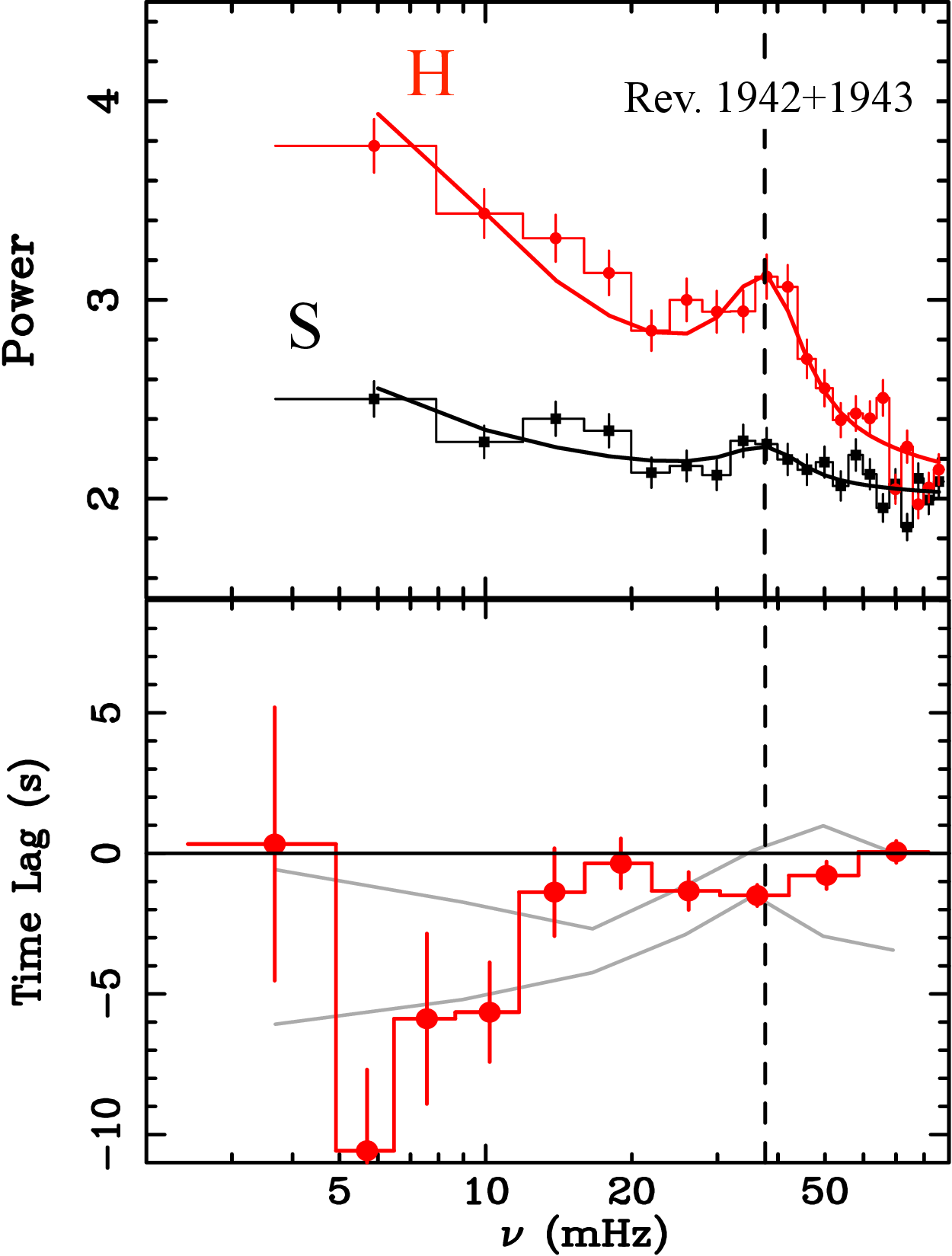} & \includegraphics[height=8.cm,width=6.8cm,angle=0]{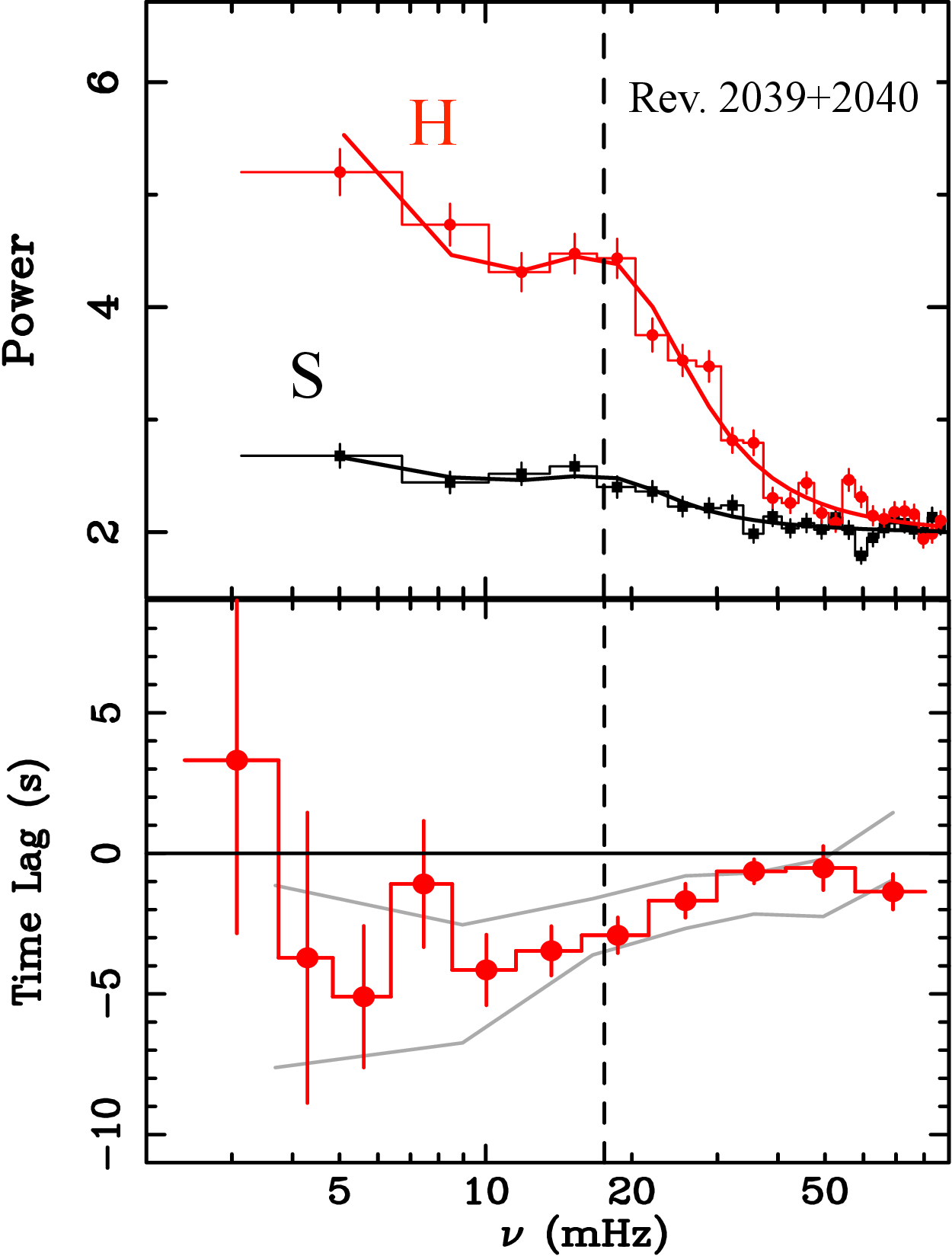} 
\end{tabular}
\caption{\emph{Upper panels:} PSD in the soft (0.3-1 keV, black) and hard (1-7 keV, red) bands. The best fit model is overplotted on the data. The vertical dashed line marks the best fit QPO centroid frequency. \emph{Lower panels:} Time lag vs frequency spectra as computed between the soft and hard energy bands. Overplotted are the 2$\sigma$ contour plots obtained from the average lag vs frequency spectrum of all the remaining observations (light grey curves).}
\label{lag_pow}

\end{figure*}

The recent analysis by PS12 of the entire set of XMM-Newton observations of NGC5408 X-1 studied in this paper, revealed significant variability in the properties of the mHz QPO (i.e. centroid frequency, \emph{rms}). 
We show in Fig. \ref{lag_pow} (upper panels) the PSD of all the observations in the soft, 0.3-1 keV, and hard, 1-7 keV, energy bands, adopting the Leahy et al (1983) normalization, whereby the Poisson noise component has constant power $P_{Poiss}=2$.
As for the fractional \emph{rms} spectrum, we combined adjacent observations during which the source is consistent with being stationary (Sect. \ref{sect:rms}). It is worth noting that, according to PS12, these also correspond to observations between which the QPO properties do not change significantly.\\
We fit the PSD with simple models using either a broken power law or broad Lorentzian components for the continuum (both the choices give similar results), and one narrow Lorentzian for the fit of the QPO.\\
There are indications of a flattening of the PSD at low frequencies, but deriving strong constraints on the presence of a break is not possible, due to limited low-frequency sampling.
A QPO is observed during all the observations, in agreement with previous results (S07, SM09, PS12). The best fit values for the QPO centroid frequency, $\nu_{QPO}$, have been marked with a vertical dotted line in the plots of Fig. \ref{lag_pow}, and are consistent with those derived by PS12. The QPO is detected with high confidence in the hard band (null hypothesis $F-$test probabilities $>10^{-10}$), with the exception of Rev. 1483, during which the significance is slightly lower (null hypothesis $F-$test probability $\sim 5\times 10^{-3}$). The best fit full-width-half-maximum (\emph{FWHM}) of the QPO in the hard band ranges between $FWHM\sim16-21$ mHz during all but Rev. 1483, where the QPO results narrower ($FWHM\sim 1$ mHz). 
On the other hand the QPO in the soft band is much weaker during Rev. 1117 and Revs. 2039+2040 (respectively a factor $\sim$3-7 weaker than the hard band QPO, and corresponding null hypothesis $F-$test probabilities $\sim 2\times 10^{-3}$), while it results undetected during the remaining observations.
The QPO quality factor (defined as $\nu_{QPO}/FWHM$) is generally relatively low, ranging between $\sim 1-3$, with the exception of Rev. 1483, where the quality factor rises to $\sim$10. These low values may be the consequence of close QPO pairs being smeared out in the rebinning process. \\
In Fig. \ref{fig:qpo_freq} we show the best fit centroid frequency values of the QPO detected in NGC 5408 X-1 as a function of the $1-50$ mHz fractional \emph{rms} in the hard band. We observe that the frequency of the QPO appears to decrease as the fractional \emph{rms} increases. The frequency of the QPO is the highest during the observations with the lowest level of hard X-ray variability (i.e. Revs. 1942+1943, see Figs. \ref{fig:stat}, \ref{fig:rms_cr}, and \ref{fig:rms_spec}). According to the classification of Mu\~noz-Darias et al (2011) for BH X-ray binaries, lower variability is observed when the source approaches softer states. Thus, provided the same classification applies for ULXs, the frequency of the QPO tends to increase when the source is softer. 
All these properties closely resemble those characterizing the well-studied type-C QPOs in X-ray binaries, which are commonly observed during both hard and intermediate states (e.g. Motta et al 2011). 

\subsection{Time lags}
\label{Sect:tlags}

Following prescriptions by Nowak et al (1999) we computed time lags in the Fourier frequency domain.
We extracted light curves in the previously defined soft and hard energy bands. Then we measured cross-spectra for consecutive light curve segments of length 1000 s, averaged them and rebinned over frequencies adopting bins of length $\nu\rightarrow 1.4\nu$. 
The phase of the average cross-spectra was finally used to derive phase lags and time lags. 
As for PSDs, we combined data from Revs. 1942-1943, and Revs. 2039-2040, while computing separate lag spectra for the others.
Fig. \ref{lag_pow} (lower panels) shows the time lag spectra as a function of frequency. 
A significant drop ($>>99.9$ percent significance from a simple $\chi^2$ test against a constant, zero-lag model) to negative amplitudes is always detected at frequencies $\nu \gsim 5$ mHz, thus indicating the presence of a soft X-ray band time delay (of the order of few seconds) with respect to the hard band.
To check for variability in the lag profile we took the difference between each lag vs frequency spectrum (Fig. \ref{lag_pow}, lower panels). For all of the 6 possible combinations we carried out a $\chi^2$ goodness-of-fit test against a constant model fixed at zero. As for the test of stationarity, we set at 0.1 percent the confidence level for rejection of the null hypothesis, i.e. that each difference lag profile is consistent with zero over the entire 1-50 mHz frequency range. In all cases we do not find significant deviations from the constant model, thus concluding that the measured time lag is statistically consistent with being unchanged over the different observations. Overplotted on each time lag profile in Fig. \ref{lag_pow} (lower panels) are the 2$\sigma$ contour plots as obtained from the average lag profile of all the remaining observations.\\
Since the time lag detected in NGC 5408 X-1 does not show any significant variation, despite the non-stationarity of the dataset (Sect. \ref{sect:stat}), we combined all the data to obtain an average lag profile.
The average phase/time lag vs frequency spectrum from all the available data is shown in Fig. \ref{lag_freq_all}. It is consistent with the existence of a constant phase shift over frequencies $\gsim 5$ mHz, corresponding to a soft time delay whose absolute amplitude decreases with frequency. The maximum soft time delay of $\tau \sim 5$ s is oberved at $\nu\sim 5-20$ mHz, while at higher frequencies ($\nu\gsim 20$ mHz) the amplitude of the lag decreases to $\tau \sim 1$ s.\\
We used the combined data to study the energy dependence of the lag in these two frequency ranges.
The lag-energy spectrum was obtained by computing cross-spectra between consecutive energy channels and the entire E$=$0.3-7 keV range as reference band, so as to increase the cross-spectra S/N ratio. To avoid contribution from correlated Poisson noise, each energy channel was excluded from the reference band before computing the corresponding cross-spectrum (as described in Zoghbi, Uttley \& Fabian 2011). A lag measurement was then derived, by averaging the resulting cross spectra over the frequency intervals $\nu\sim 5-20$ mHz and $\nu\sim 20-90$ mHz. 
 The result is shown in Fig. \ref{lag_E}, where all the points have been shifted so that the largest (in absolute value) time lag value marks the leading energy band. Softer energy bands always lag harder ones and the amplitude of the time lag increases as the energy increases, with a maximum time delay between the softest and the hardest energies of $\tau \sim 5$ s in the low frequency interval and $\tau \sim 1.5$ s in the high frequency interval.

\begin{table*}
\caption{Summary of the spectral/variability properties analyzed for NGC 5408 X-1 as a function of accretion state in X-ray binaries. Text in boldface refers to the observed properties of the ULX. }
\label{summ}
\centering
\vspace{0.2cm}
\begin{footnotesize}
\begin{tabular}{c|c|c|c|c}
\hline\hline
               &  LHS & HIMS & SIMS & HSS  \\
\hline
\vspace{0.2cm}
Disc fraction$^a$ & $\lsim$ 20 percent&  \multicolumn{2}{c}{{\bf $\lsim$ 80 percent}} &  $\gsim$ 75 percent \\
\vspace{0.2cm}
\emph{rms} vs count rate$^b$ & linear (low scatter) &  \multicolumn{2}{c}{{\bf chaotic (overall \emph{rms} decreases as the flux increases)}} &  linear (high scatter) \\
\vspace{0.2cm}
fractional \emph{rms}$^b$ & {\bf $\sim$30-50 percent} & {\bf $\lsim$30 percent} & $\lsim$ 10 percent & $\lsim$ 5 percent\\
fractional \emph{rms} vs E$^c$ & flat/decreasing & flat/decreasing (early stages) or & {\bf increasing}  & {\bf increasing}  \\
\vspace{0.2cm}
                       &                    &           {\bf increasing (late stages)}         &    &  \\
QPO$^{d,e}$     & {\bf type C} & {\bf type C}  & type A/type B & no \\

\hline
\hline
\end{tabular}
\\
References: $^{(a)}$ Remillard \& McClintock (2006); $^{(b)}$ Mu\~noz-Darias et al (2011); $^{(c)}$ Belloni et al (2011); $^{(d)}$ Casella et al (2004); $^{(e)}$ Motta et al (2011).
\\
\end{footnotesize}
\end{table*}

\section{Discussion}
\label{discus}
In this paper we carried out a comprehensive timing analysis of all the archived XMM-Newton observations (public as of February 2012) of the ULX NGC 5408 X-1. 
The short time scale variability ($\nu=1-50$ mHz) of the source in the hard energy band (1-7 keV) shows significant deviations from stationarity among different observations. On the other hand, the variability in the soft band (0.3-1 keV) is very low, and the relative errors do not allow us to detect any significant variation among the observations (see Sect. \ref{sect:stat} and Fig. \ref{fig:stat}).

\subsection{Association with accretion states of BHBs}
\begin{figure} 
\centering

\begin{tabular}{p{4.4cm}}
\includegraphics[height=5.8cm,angle=270]{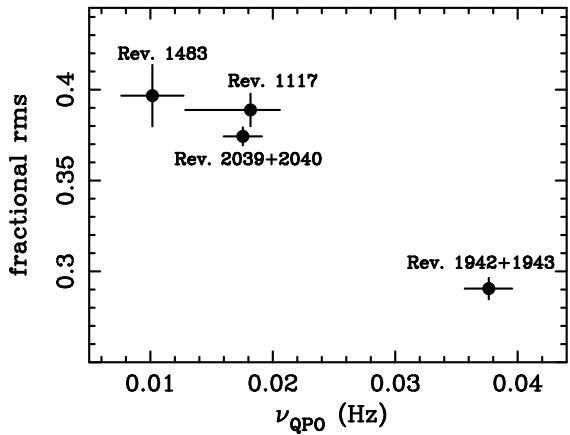} 
\end{tabular}
\caption{The best fit centroid frequency of the QPO as a function of 1-50 mHz fractional \emph{rms} in the hard (1-7 keV) energy band.}
\label{fig:qpo_freq}

\end{figure}

\begin{figure}
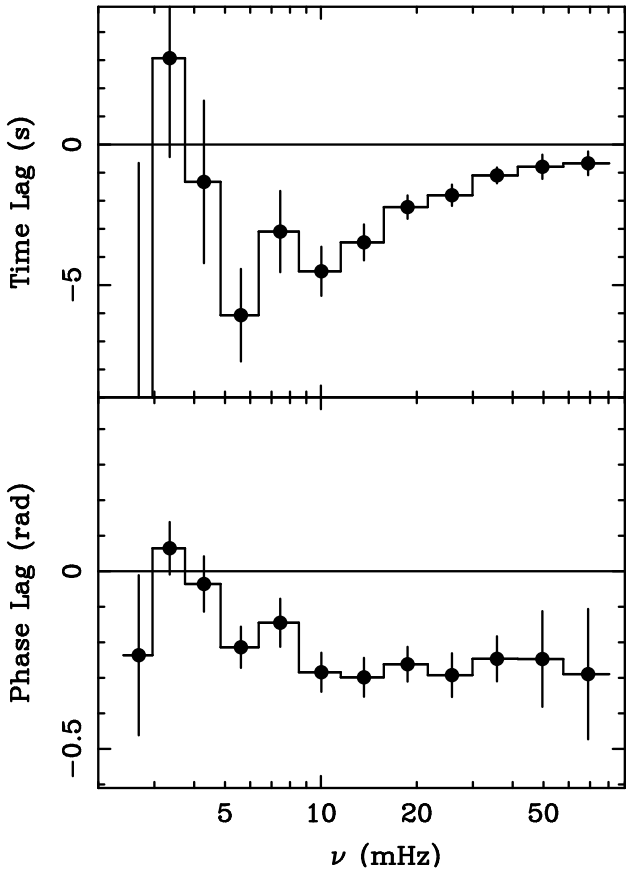

\centering

\includegraphics[angle=270,scale=0.35]{figure/tl_all.ps}\\
\vspace{-1.36cm}
\includegraphics[angle=270,scale=0.35]{figure/ph_all.ps}

  \caption{Time lag (\emph{upper panel}) and corresponding phase lag (\emph{lower panel}) as a function of frequency as computed between energy bands $E=0.3-1$ keV and $E=1-7$ keV averaging over all the archived observations of NGC 5408 X-1.}
\label{lag_freq_all}
    \end{figure}

From the point of view of spectral components, the analyzed energy bands are dominated by a hard X-ray curvature with a turn over at $E\gsim$ 4-6 keV, and a soft X-ray excess (e.g. Gladstone et al 2009). According to previous spectral analyses of NGC 5408 X-1 (S07, SM09, Gladstone et al 2009, PS12) the two components can be well fitted by a thermal disc (soft component) and a Comptonization model from a warm and thick corona (hard component). While a soft excess associated to a thermal disc component is commonly observed during some X-ray binary accretion states, the hard break is always detected at much higher energies (E$\gsim$30 keV depending on the accretion state of the source, e.g. Motta et al. 2009, Del Santo et al 2008). We assumed a phenomenological model comprising contribution from multi-colour disc-blackbody emission plus Comptonization, and including absorption by a column of cold gas (\emph{phabs$*$simpl$*$diskbb} in Xspec, Steiner et al 2009). We fitted all the XMM-Newton spectra of NGC 5408 X-1 and computed the unabsorbed disc flux fraction ($L_{Disc}/(L_{Disc}+L_{PL})$, see Dunn et al 2011) in the 0.3-10 keV energy band. We found that this ratio is always in the range $\sim$0.61-0.68, which would agree with the source being in an intermediate state (e.g. Remillard \& McClintock 2006; Dunn et al 2011; see Table \ref{summ}, where the properties analyzed in this paper have been classified as a function of accretion state in X-ray binaries).\\
Independent indications about the accretion state of the source come from its variability properties (see Table \ref{summ}). 
We found that the observed \emph{rms} variations in the hard energy band are not linearly correlated with the hard X-ray count rate (Sect. \ref{sect:stat} and Fig. \ref{fig:rms_cr}), excluding that the source is in an accretion state analogous to the canonical LHS of BH X-ray binaries (Mu\~noz-Darias et al 2011).\\
The short time scale ($\nu=1-50$ mHz), fractional \emph{rms} spectrum of NGC 5408 X-1 shows a spectral hardening (already reported by Middleton et al 2011 for Revs. 1117 and 1483) during all the XMM-Newton observations (see Sect. \ref{sect:rms} and Fig. \ref{fig:rms_spec}), with the hard band reaching quite high variability levels (up to $\sim$ 30 percent of fractional \emph{rms}). 
The amount of variability as a function of energy is observed to vary during BH X-ray binary outbursts (Belloni et al 2011).
In particular, the \emph{rms} spectrum, which is mostly flat during LHS, undergoes major changes when the source moves to softer states. At the softest end of HIMS a sudden spectral hardening of the variability is observed, with the overall normalization of the \emph{rms} spectrum decreasing towards softer accretion regimes, i.e. through the SIMS up to the canonical disc-dominated HSS. 
This behaviour is similar to what is observed in NGC 5408 X-1. The latter property plus the detected high levels of variability lead us to the conclusion that the source is most probably observed during a HIMS.\\
As previously reported by S07, SM09, and PS12, a mHz QPO is observed during all the observations. The QPO is detected with high significance in the hard band, and the best fit centroid frequency is variable among the different observations, shifting towards higher frequencies when the fractional \emph{rms} decreases (see Fig. \ref{fig:qpo_freq}). 
According to Mu\~noz-Darias et al (2011), the decrease of \emph{rms} is indicative of a softening of the source.
Unfortunately, the number of observations and the range of spanned values of fractional \emph{rms} is small (the largest variation observed is a factor of $\sim 1.3$) to test for the presence of a correlation between the QPO frequency and the fractional \emph{rms}. However, the observed trend is in the sense of an increase of QPO frequency as the source becomes softer\footnote{The correlation between QPO frequency and spectral index has been investigated in PS12. The authors found that the range of spectral variations is modest (about a factor of 1.1, thus of the same order of the fractional \emph{rms} variation reported in this paper) as compared to the variations in the QPO centroid frequency. This behavior is typical of X-ray binaries in intermediate states, during which the type-C QPO frequency vs spectral index correlation appears to saturate (Vignarca et al 2003; Shaposhnikov \& Titarchuk 2009). This conclusion is in agreement with our results from timing analysis, and would indicate that the source is persistently observed in an intermediate state over a time interval of about 5 years.}.
This is qualitatively consistent with the behavior characteristic of type-C QPOs (e.g. Reig et al 2000). This kind of QPOs are commonly observed during intermediate states. 
It is worth noting that type-C QPOs are observed in association with a flat-topped PSD and high total variability level. 
Although strong constraints on the shape of the broad-band PSD cannot be derived due to limited low-frequency sampling, the observed high variability level is in agreement with the type-C QPO identification.
However, as pointed out by PS12 and Middleton et al (2011), the QPO in NGC 5408 X-1 also shows some properties that do not easily reconcile with standard type-C QPOs as observed in X-ray binaries, e.g. a low quality factor (see Sect. \ref{Sect:PSD}) and the lack of a correlation with the PSD break. It is worth noting that a higher frequency resolution would allow to test for the presence of multiple harmonic peaks which might result smeared out in the rebinning of current data, thus decreasing the observed quality factor, while a better low frequency coverage\footnote{Note that the problem of poor low frequency coverage can be circumvented by the use of simulations (e.g. Middleton et al 2011).} would confirm the lack of a correlation with the PSD break.

\subsection{Soft lag origin}

We report the highly significant (at $>>99.9$ percent level) detection of a soft (negative) lag of $\sim$5 s at its maximum (in absolute value) amplitude in the X-ray light curves of the source, spanning 
the range of frequencies $\nu\sim 5-90$ mHz. The lag profile is consistent with being the same over all the analyzed observations (see Fig. \ref{lag_pow}, lower panels), despite the non-stationarity of the dataset (Sect. \ref{sect:stat}).
 It is worth noting that the detection of a soft lag in this source was previously reported by Heil \& Vaughan (2010), from the analysis of
Revs. 1117 and 1483 only. The addition of the 2010 and 2011 observations allowed us to better constrain the lag profile, and also to study its energy dependence. 
We find that the phase/amplitude (in absolute value) of the lag increases towards harder energy bands (Fig. \ref{lag_E}).
In the following we discuss analogies with soft X-ray lags detected in other accreting BH systems, and possible interpretations.

\subsubsection{Lag-QPO association}
\label{sect:qpo}
The frequency range where the soft lag is detected encompasses the QPO frequencies in NGC 5408 X-1 ( Fig. \ref{lag_pow}). 
Soft X-ray lags displaying an association with QPOs and their harmonics (we will refer to them as QPO lags) have been detected in a number of BH X-ray binaries (e.g. XTE J1550-564 Remillard et al 2002; XTE J1859+226 Casella et al 2004; GRS 1915+105 Reig et al 2000), and in a couple of neutron stars (de Avellar et al 2013), although it is worth noting that these results are mostly obtained from RXTE data, which is not sensitive in the EPIC-pn soft X-ray band, thus precluding the study of QPO lags at these energies.\\
Interestingly, the properties of QPO lags show a dependence on the QPO type. Although the majority are characterized by a hard/positive delay at the fundamental QPO frequency (Casella et al 2004), those associated with type-C QPOs can switch to soft/negative values (e.g. Reig et al 2000).
Indeed, in a couple of sources the amplitude of type-C QPO lags has been observed to smoothly decrease with increasing QPO frequency and to change from positive to negative (i.e. from hard to soft delays) above $\nu_{QPO}\sim$2 Hz (GRS 1915+105, Reig et al 2000; XTE J1550-564 Remillard et al 2002). As discussed in the previous Section, high frequency type-C QPOs are observed during the softest stages of intermediate states (e.g. Reig et al 2000, Casella et al 2004). Thus, provided the QPO-lag association holds, the observation of a soft lag is naturally expected in NGC 5408 X-1.\\
It is worth noting that a similar phase lag trend has been observed also in the continuum around the QPO frequency (Reig et al 2000).\\
Consequently, a possible origin for the soft X-ray lag in NGC 5408 X-1 may involve an association with the QPO observed in its PSD. 
Besides, the lag energy dependence reported in this paper (Fig. \ref{lag_E}) is reminiscent of the trend observed in the energy spectra of type-C QPO soft lags (Fig. 4 of Reig et al 2000 for the case of GRS 1915+105).
Hence, the properties of the soft lag in NGC 5408 X-1 are in agreement with those characterizing lags associated to type-C QPOs in BHBs. A comparison between the QPO frequency of NGC 5408 X-1 ($\nu \sim 10-38$ mHz) and the range of frequencies of type-C QPOs at which soft lags are observed  ($\nu\sim2-8$ Hz), yields a scaling factor of $\sim 200$ for the BH mass. The same linear scaling factor is implied when comparing the lag amplitudes. For a typical binary system hosting a BH of tens of solar masses, this would correspond to a rescaled BH mass of the order of $\gsim 10^3$M$_{\odot}$ for NGC 5408 X-1, consistent with estimates obtained through other methods (e.g. Casella et al 2008).\\
Nonetheless, an identification of the QPO observed in NGC 5408 X-1 with the mHz QPOs observed in BHBs during bright states (Morgan et al 1997) cannot be completely excluded based on these results. Indeed Cui (1999) demonstrated that the lags associated with mHz QPOs in GRS 1915+105 have a complex behaviour, occasionally resembling the trend we observe in the lag-E spectrum of NGC 5408 X-1 (Fig. \ref{lag_E}). This interpretation would imply super-Eddington accretion onto a $\leq$100 M$_{\odot}$ BH (Middleton et al 2011).\\
However, some discrepancies with the QPO-lag association emerge from our analysis. The QPO-like feature in NGC 5408 X-1 is significantly weaker or undetected in the soft band ($E<1$keV, see Fig. \ref{lag_pow}), in agreement with S07 and SM09. Conversely, a soft lag (Fig. \ref{lag_E}) is significantly detected even at energies $E\lsim 1$ keV.
Moreover, the soft lag in NGC 5408 X-1 extends over a relatively broad range of frequencies $\nu \sim 5-90$ mHz, a factor of $\sim2$ broader than the width of the Lorentzian used to fit the QPO. 
The large error bars at the lower frequency end, and the predominance of Poisson noise at the high frequency end, do not allow to rule out an even broader profile.
Finally, contrary to what observed (Sect. \ref{Sect:tlags}) the lags should change as the QPO varies.\\
These observational evidences seem to suggest that the delayed response is not only in the QPO component, but rather involves a larger part of the variability continuum, i.e. on a larger range of variability time scales (as previously noted, delays in the continuum around the QPO have been observed in BHBs as well, e.g. Reig et al 2000).

\subsubsection{Reverberation lags}

High frequency, soft X-ray lags have been observed in several accreting systems, and interpreted as the signature of reverberation (e.g. Fabian et al 2009, Zoghbi et al 2010, Emmanoulopoulos et al 2011, Uttley et al 2011). It is thus interesting to test whether the soft lag in NGC 5408 X-1 can have the same origin.
In Fig. \ref{corr_plot} we report the observed correlation between soft X-ray lags amplitude and BH mass in radio quiet AGN (De Marco et al 2013), which has been so far tested over the range $\sim 10^{5-9}$M$_{\odot}$ of BH masses. In the low mass-end of the plot we added the three points that represent detections of soft X-ray lags in the BHB GX 339-4 (Miyamoto et al 1993, Uttley et al 2011), and in the two neutron star X-ray binaries 4U 1608-52 and 4U 1636-53 (de Avellar et al 2013).\\
The few milliseconds soft lag in GX 339-4 has been observed in both a hard state (Uttley et al 2011) and an intermediate state (Miyamoto et al 1993) of the source, meaning that, if interpreted as thermal reverberation off the accretion disc one must take into account the fact that the disc might be truncated (Kubota \& Done 2004, Uttley et al 2011, Plant et al 2013). 
On the other hand, the soft X-ray lags of few $\mu$s recently reported in the neutron-star X-ray binaries 4U 1608-52 and 4U 1636-53 are observed at the frequencies of the kHz QPOs, which might be the signature of Keplerian modulation at the inner edge of the accretion disc (e.g. Stella \& Vietri 1999, McClintock \& Remillard 2006 and references therein). One of the possible scenarios that have been proposed for these lags is that they represent the light travel time between the corona and/or the neutron star surface or boundary layer, and the innermost accretion disc (de Avellar et al 2013).\\
\begin{figure}
\centering

\includegraphics[angle=270,scale=0.33]{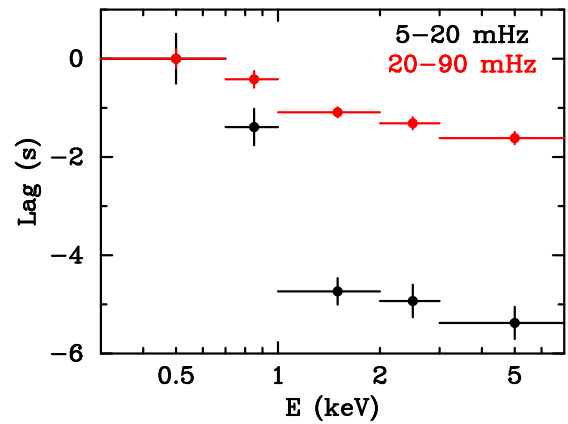}

  \caption{Time lag (in the frequency ranges $\nu=5-20$ mHz and $\nu=20-90$ mHz) as a function of energy.}
\label{lag_E}
    \end{figure}
In Fig. \ref{corr_plot} (dashed coloured lines) we overplotted the expected light crossing time trends at different gravitational radii (where the light crossing time is defined as $t_c=r_g/c=GM/c^3$). We note that the distribution of AGN data points in the plot is clustered within the range of $\sim$1-10 r$_g$ light crossing time, while the lag of GX 339-4 is rather consistent with the light crossing time at larger distances (i.e.  of the order of $\sim$50 r$_g/c$). Assuming a standard accretion disc-corona model (e.g. Haardt \& Maraschi 1991) with a geometrically thin and optically thick disc (Shakura \& Sunyaev 1973), this suggests the disc to extend down to very short distances from the central BH in the case of radio-quiet AGN and the two neutron stars X-ray binaries, while it appears to be truncated at larger distances in the case of GX 339-4 (as suggested by Uttley et al 2011), consistent with the latter source being in the canonical LHS.\\ 
Being the BH mass and disc inner radius of NGC 5408 X-1 unknown we cannot infer the position of this ULX in the plot. However, 
we note that, for the source to be in an intermediate state (either along the hard-to-soft or the soft-to-hard transition branch) the luminosity should be $L> 0.01  L_{Edd}$ (Maccarone 2003). This places a constraint on the BH mass of $\lsim 3\times 10^{4}$ M$_{\odot}$ (where we assumed $L_{Bol}\sim L_{0.3-10 keV} \sim 3.9 \times 10^{40}$ erg s$^{-1}$, as derived from the \emph{simpl$*$diskbb} fit, see Sect. \ref{discus}). On the other hand, ULXs may be sources accreting close or above the Eddington limit. In this case NGC 5408 X-1 would plausibly have a mass of few solar masses, i.e. similar to the mass of standard stellar mass accreting BHs. Thus we assume a rough lower limit of $\gsim 10$ M$_{\odot}$.  
We report in Fig. \ref{corr_plot} (yellow-shaded  box) this range of mass values. This is plotted against the observed range soft time lags (see Fig. \ref{lag_E}) in both the low and high frequency range (i.e. $\tau\sim 1.5-5.5$ s). According to this plot, the observed soft lag represents distances of the order of tens-to-hundreds of r$_g$ for an IMBH of $10^{3-4}$ M$_{\odot}$, and distances of $\gsim 1000$ r$_g$ if the BH mass is of the order of $10-100$  M$_{\odot}$. If the lag is the signature of reverberation from the accretion disc, then the former hypothesis is more reliable, and in agreement with observations of disc truncation during intermediate states (e.g. Kubota \& Done 2004, Plant et al 2013).
In this case, assuming the source is caught right during the soft-to-hard transition (thus $L\sim 0.01-0.03 L_{Edd}$, Maccarone 2003), the BH mass would be of the order of $\sim10^{4}$ M$_{\odot}$.\\
On the other hand, for a stellar mass BH, the observed lag amplitudes suggest the reprocessing medium to be at much higher distances, and/or to be much more extended, like for example in the case of a wind (e.g. Gladstone et al 2009; Middleton et al 2011). 
The emergence of an outflow within a given radius in super-critically accreting BHs has been proposed to explain the spectral properties of ULXs (e.g. Poutanen et al 2007) as well as the presence of photoionized nebulae observed around some of them (e.g. Kaaret et al 2004). In this case, hard X-rays from the inner regions would be seen through the outflow, with the soft X-ray spectrum being due to emission in the wind. However, the implications of such a model on the lag spectrum strongly depend on the geometry and properties of the wind, and are hard to predict here.

\begin{figure}
\centering

\includegraphics[angle=0,scale=1.18]{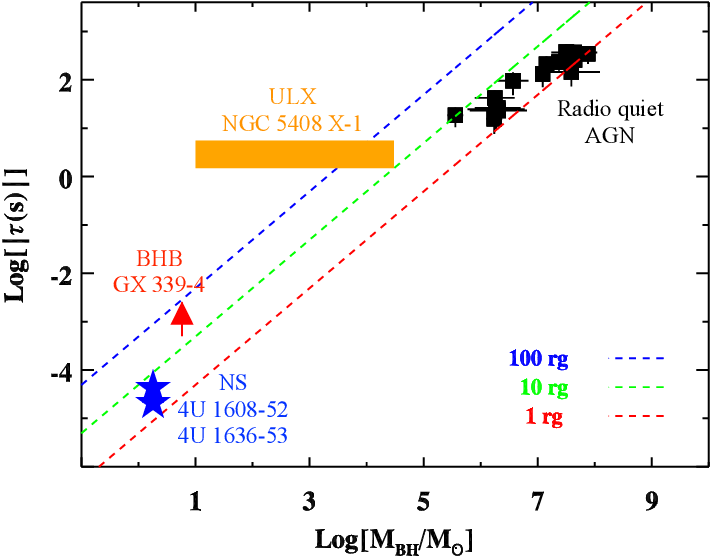}

  \caption{Correlation between soft X-ray time lag and BH mass (black lines) for the radio-quiet AGN sample (black squares). The red triangle is the soft lag observed in GX 339-4 (Uttley et al 2011), while the blue stars are the soft lags observed in the two neutron star X-ray binaries 4U 1608-52 and 4U 1636-53 (de Avellar et al 2013). The yellow box marks the possible range of mass values for NGC 5408 X-1 at the level of the detected soft lags. The dotted diagonal lines mark the light crossing time at $1 r_g$, $10 r_g$, $100 r_g$. The mass used for GX 339-4 is taken from Mu\~noz-Darias et al (2008), while for the neutron stars we assumed the range of masses $\sim 1.4-2.2$M$_{\odot}$.}
\label{corr_plot}
    \end{figure}

\section{Conclusions}
 
 We have carried out a comprehensive analysis of the fast X-ray variability and time lags of the ULX NGC 5408 X-1. Our analysis showed that:

\begin{itemize}

\item The source is consistent with being non-stationary over the time interval covered by the analyzed observations ($\sim$5 years). 

\item All the observed variability properties closely resemble those observed during the hard-intermediate state of X-ray binaries. We note that most of the known X-ray binaries undergo outbursts that last for a typical period of some months, after which they return to quiescence. The fact that NGC 5408 X-1 is observed in an intermediate state over a time interval of at least 5 years may reflect the presence of a BH with higher mass (but note, for example, the peculiar case of the stellar mass BH X-ray binary, GRS 1915+105, which has been observed in a bright state since its discovery, e.g. Fender \& Belloni 2004).

\item Confirming results by Heil \& Vaughan (2010) we report the highly significant detection of a soft lag of few seconds, in the frequency range $\sim5-90$ mHz, whereby variations in the soft X-ray band, 0.3-1 keV, lag behind those in the hard X-ray band, 1-7 keV.

\item Provided the lag is associated with the QPO, it displays characteristic properties of soft lags observed at the frequency of type-C QPOs in BHBs. This would imply a scaling factor of $\sim 200$ on the BH mass, thus favouring an IMBH interpretation.

\item If the lag is associated with the variations in the continuum, rather than with the QPO component only, it might be either the signature of reverberation off the inner regions of the accretion disc or of reprocessing in a larger scale structure, like a wind.

\item The measured lag implies a distance of the primary-to-reprocessed emission regions of 10-100 r$_{g}$ if the central object is an IMBH accreting below the Eddington limit (disc reverberation). Assuming the source is caught during a soft-to-hard transition ($L\sim 0.01-0.03 L_{Edd}$, Maccarone 2003), then the implied value of the BH mass is $\sim 10^{4}$ M$_{\odot}$.

\item The measured lag corresponds to a distance of $\gsim$ 1000 r$_{g}$ for a $10-100$ M$_{\odot}$ BH accreting at high rate (reprocessing in a wind). 

\end{itemize}
If the analogy with soft X-ray lags observed in AGN (De Marco et al 2013), in the BHB GX 339-4 (Uttley et al 2011) and in the two neutron star X-ray binaries 4U 1608-52 and 4U 1636-53 (deAvellar et al 2013) holds, then  the hypothesis of an IMBH seems more coherent with our overall results, and implies a truncation radius of the order of tens-to-hundreds of r$_g$. In this case, the observed soft lag in NGC 5408 X-1 would fit in an unified picture, which links the soft X-ray lag properties in sources spanning about 8 orders of magnitude in BH mass.\\
It is interesting to note that intensive \emph{Swift} monitoring of the source suggests the presence of dips in the X-ray light curves (Pasham \& Strohmayer 2013, Gris\'e et al 2013). The nature of these dips is not yet clear, but it might imply that the system is observed almost edge-on. High inclination BH X-ray binaries have been demonstrated to show higher observed inner disc temperatures than low inclination systems, i.e. $T_{obs}\gsim 0.6$ keV (Mu\~noz-Darias et al 2013). Our best fit values of the inner disc temperature (i.e. $\sim 0.16$ keV, as obtained from the fit with the \emph{phabs$*$simpl$*$diskbb} model) are a factor $\gsim$ 4 lower (although note that Gladstone et al 2009 argue that this observed cool temperature might not be associated with direct emission from the disc), thus implying a limit of M$_{BH}\gsim 10^3$ M$_{\odot}$ for the BH mass (e.g. Miller et al 2004), in line with the IMBH interpretation.\\
It is however important to remark that, despite the variability properties of the source are well in agreement with those characterizing X-ray binaries during HIMS, the spectral properties remain puzzling. In particular, in our computation of the disc fraction we discarded the presence of a turn over at E$\gsim$ 4-6 keV, since it does not affect significantly the analyzed range of energies. However this feature may represent the signature of a warm, optically thick corona (e.g. Gladstone et al 2009), which is a peculiar feature of some ULXs and is rarely observed in X-ray binaries (e.g. Ueda, Yamaoka \& Remillard 2009). Although standard models (e.g. reflection, Caballero-Garc\'{i}a \& Fabian 2010) have been shown to fit equally well the data, the correct interpretation is still a matter of debate.

\section*{Acknowledgments}

This work is based on observations obtained with XMM-{\it Newton}, an ESA science mission with instruments and contributions directly funded by ESA Member States and NASA. BDM and GM thank the Spanish Ministry of Science and Innovation (now Ministerio de Econom\'ia y Competitividad) for financial support through grant AYA2010-21490-C02-02. GM also thanks the European Union Seventh Framework Program (FP7/2007--2013) for support under grant 312789. TB acknowledges support from the EC Seventh Framework Programme (FP7/2007-2013) under grant agreement number ITN 215212 ``Black Hole universe'' and from the Leverhulme Trust through a LT Visiting Professorship.
GP acknowledge support via an EU Marie Curie Intra-European fellowship under contract no. FP-PEOPLE-2012-IEF-331095. The authors thank the anonymous referee for helpful comments which allowed to significantly improve the paper.


\end{document}